\documentclass[aps,prl,8pt,twocolumn,showpacs,amsmath,amssymb,floatfix,footinbib,superscriptaddress]{revtex4-1}
\usepackage{graphics,graphicx,epsfig,ulem,epstopdf,bm,longtable,url,geometry,datetime}
\usepackage[dvipdfm,colorlinks=true,linkcolor=blue,urlcolor=blue,citecolor=blue]{hyperref}
\usepackage{amsmath,amssymb,latexsym,float,amsfonts,subfigure,hyperref,color}
\usepackage[ansinew]{inputenc}
\usepackage[usenames,dvipsnames]{pstricks}
\usepackage[pagewise,mathlines]{lineno}
\def\footnoterule{\kern -1mm \hrule width 6.0cm \kern 2.2mm}%
\usepackage{tikz,xcolor,hyperref}
\definecolor{lime}{HTML}{A6CE39}
\DeclareRobustCommand{\orcidicon}{%
    \begin{tikzpicture}
    \draw[lime, fill=lime] (0,0)
    circle [radius=0.16]
    node[white] {{\fontfamily{qag}\selectfont \tiny ID}};\draw[white, fill=white] (-0.0625,0.095)
    circle [radius=0.007];
    \end{tikzpicture}
    \hspace{-2mm}}
\foreach \x in {A, ..., Z}
{\expandafter\xdef\csname orcid\x\endcsname{\noexpand\href{https://orcid.org/\csname orcidauthor\x\endcsname}{\noexpand\orcidicon}}}

\begin{document}
\title{Non-Markovian N-spin chain quantum battery in thermal charging process}

\author{Shun-Cai Zhao\orcidA{}}%
\email[Corresponding author: ]{zhaosc@kust.edu.cn.}
\affiliation{Center for Quantum Materials and Computational Condensed Matter Physics, Kunming University of Science and Technology, Kunming, 650500, PR China}
\affiliation{School of Science, Department of Physics, Kunming University of Science and Technology, Kunming, 650093, PR China}

\author{Zi-Ran Zhao}
\email[Co-first author.]{zscgroup@126.com}
\affiliation{Center for Quantum Materials and Computational Condensed Matter Physics, Kunming University of Science and Technology, Kunming, 650500, PR China}
\affiliation{School of Science, Department of Physics, Kunming University of Science and Technology, Kunming, 650093, PR China}

\author{Ni-Ya Zhuang}
\affiliation{Center for Quantum Materials and Computational Condensed Matter Physics,  Kunming University of Science and Technology, Kunming, 650500, PR China}
\affiliation{School of Science, Department of Physics, Kunming University of Science and Technology, Kunming, 650093, PR China}

\date{ \today}
\begin{abstract}
Ergotropy serves as a key indicator for assessing the performance of quantum batteries(QBs). Using the Redfield master equation, we investigate ergotropy dynamics in a non-Markovian QB composed of an N-spin chain embedded in a microcavity. Distinct from Markovian charging process, the thermal charging process exhibits a distinct oscillatory behavior in the extracted ergotropy. We show that these oscillations can be effectively suppressed through coordinated tuning of coherent driving, cavity parameters, and spin-spin couplings. In addition, we analyze the influence of various system and environmental parameters on the time evolution of ergotropy, revealing rich dynamical features. Our results offer new insights into the control of energy extraction in QBs and may inform future designs of practical battery architectures.
\end{abstract}

\maketitle
\tableofcontents
\section{Introduction}\label{Introduction}

The ability to store and extract useful work from quantum systems\cite{PhysRevLett.120.117702,PhysRevA.97.022106,PhysRevE.99.052106} has attracted significant interest in the emerging field of quantum thermodynamics\cite{Vinjanampathy2016ContPhys,Benenti2017PhysRep}. Among various proposals, quantum batteries (QBs) that store energy and deliver it in the form of work-have been widely studied due to their potential in nanoscale energy storage and quantum technologies\cite{PhysRevLett.111.240401,Binder_2015,Hovhannisyan2020PRResearch,Quach2023Joule,Campaioli2024RMP}. A key performance indicator of QBs is ergotropy, which quantifies the maximum extractable work from a quantum state via unitary operations without entropy change\cite{PhysRevE.105.L052101,PhysRevLett.125.180603,PhysRevA.109.052219,PhysRevLett.134.010408}.

Recent studies have investigated QB charging dynamics across various architectures. Classical models based on spin chains\cite{PhysRevA.103.033715,PhysRevA.97.022106,Grazi2024PRL,Catalano2024PRXQuantum}, harmonic oscillators\cite{PhysRevA.107.042419,PhysRevB.98.205423,PhysRevE.99.052106}, and cavity QED systems\cite{Hadipour202400115} have demonstrated that entanglement\cite{PhysRevA.107.022215} and collective effects\cite{PhysRevA.105.062203,PhysRevLett.118.150601} can enhance charging power and efficiency. In particular, thermal charging schemes driven by bosonic reservoirs have been proposed as physically realistic alternatives to coherent protocols\cite{PhysRevA.108.052213,PhysRevA.99.032345,PhysRevB.110.085419}. However, most existing studies adopt Markovian master equations\cite{PhysRevA.98.052129}, such as the Lindblad or Gorini-Kossakowski-Sudarshan-Lindblad formalisms, which assume weak system-environment coupling and negligible memory effects\cite{PhysRevA.109.042424,Li2022,PhysRevA.109.012224}. For instance, Ref. \cite{PhysRevA.103.033715} examined thermal charging under a Markovian setup, reporting steady-state ergotropy but neglecting memory-induced oscillations.

To address this limitation, we consider a strongly coupled, non-Markovian QB model consisting of an N-spin chain embedded in a single-mode microcavity. The Redfield master equation is adopted to describe the thermal charging process beyond the weak-coupling limit. This framework enables a more accurate exploration of the dynamical behavior of ergotropy, especially in regimes where traditional Lindblad-based treatments are insufficient. Specifically, we address how system parameters-including inter-spin hopping, cavity characteristics, and external coherent driving-can be tuned to stabilize and optimize the charging performance. These results provide deeper insights into quantum thermodynamic processes and inform the design of practical QB architectures bridging theory and technology.

Our paper is organized as following. In Sec.\ref{model}, we provide an N-spin chain QB embedded in a single-mode microcavity and discuss its thermal charging process. Sec.\ref{discussion} contains the main results we obtained, we report on the dynamics of QB thermal charging. The conclusion is given in Sec.\ref{conclusion}.

\section{Theoretical Model}\label{model}

\begin{figure}[h]
\centering
\includegraphics[width=0.8\columnwidth]{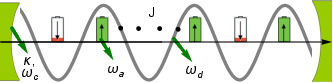}
\caption{Microcavity-based N-spin chain quantum battery scheme. \(\omega_{c}\) and \(\kappa\) denote the cavity field oscillation frequency and dissipation rate, respectively, characterizing system-environment coupling. \(\omega_{a}\) is the single-spin flip frequency, and \(\omega_{d}\) is the driving field frequency. The hopping interaction between the nearest-neighbor spin is depicted by a strength of \(J\).} \label{Fig1}
\end{figure}

To elucidate the connections and emphasize the distinctions between classical and QBs, we present a cavity-mediated spin-chain QB architecture. This work develops a robust charging protocol that addresses the critical challenge of weak charger-QB coupling in microcavity-confined systems, thereby providing crucial insights for implementing practical quantum energy storage devices. The QB is modeled by a N-spin chain as shown in Fig.(\ref{Fig1}).The total Hamiltonian of this QB system is read as follows:

\vskip -0.1cm\begin{equation}
\hat{H}_{t} =\hat{H}_{b}+\sum_{i=1}^{n}\omega_c \hat{a}_i^\dagger \hat{a}_i + \hat{H}_{\text{int}}  ,\label{eq1}
\end{equation}

\noindent In the above formalism, the system Hamiltonian \(\hat{H}_{b}\) describes the QB in Eq.(\ref{eq2}), while the interaction Hamiltonian \(\hat{H}_{\text{int}}\) incorporates both the spin-cavity coupling and the contribution from the driven field in Eq.(\ref{eq3}). They are listed as follows,

\vskip -0.1cm\begin{equation}
\hat{H}_{b}=\sum_{i=1}^{N} \omega_{a} \hat{\sigma}_i^{+} \hat{\sigma}_i^{-}+J  \sum_{i=1}^{N-1}(\hat{\sigma}_i^+\hat{\sigma}_{i+1}^- + H.C.) ,\label{eq2}
\end{equation}

\vskip -0.1cm\begin{equation}
\hat{H}_{\text{int}} = g \sum_{i=1}^{N}(\hat{a}^\dagger \hat{\sigma}_i^- + \hat{a}\hat{\sigma}_i^+) + Asin(\omega_{d}t)(\hat{a}^\dagger + \hat{a}),\label{eq3}
\end{equation}

\noindent where \(\hat{\sigma}_i^{+,-}\) denote the spin raising and lowering operators associated with the spin frequency \(\omega_{a}\) of the
N-spin chain QB. The parameter \(J\) represents the nearest-neighbor hopping interaction. The annihilation (creation) operator \(\hat{a}_i^\dagger\) \((\hat{a}_i)\) characterizes the cavity field with frequency \(\omega_{c}\). During the charging process, the initially empty QB is charged by extracting energy from the cavity field. The interaction term between the spin and the cavity field with spin-photon coupling constant \(g\) is written in Eq.(\ref{eq3}). A periodic coherent driving field with frequency \(\omega_{d}\) is included in the second term of Eq.(\ref{eq3}), representing an externally applied charging mechanism in the macroscopic regime. This time-dependent drive acts on the spin subsystem and serves as an additional energy source for the QB. It enables tunable control over energy injection and modulates ergotropy dynamics through interference with cavity-induced processes. In the absence of driving (\(A\)=0), energy transfer is governed solely by the thermal environment, providing a baseline scenario dominated by dissipative dynamics.

\begin{equation}
\frac{d\hat{\rho}}{dt} = \mathcal{L}[\hat{\rho}] = -\frac{i}{\hbar}[\hat{H}_{t}, \hat{\rho}] + \mathcal{R}_L^{(\omega_{c})}[\rho],\label{eq4}
\end{equation}
\vskip 0.5cm

The dynamic process of the N-spin chain QB thermal coherent charging is obtained by solving the Redfield master Eq.(\ref{eq4})\cite{zhao2025}. The first part of Eq.(\ref{eq4}) describes the unitary evolution of the QB system, while the second term denotes the Redfield dissipative superoperator, capturing non-unitary dissipation under the non-Markovian approximation of the Redfield master equation.
While the Redfield master equation is formally derived under the Born approximation and is thus most accurate in the weak-coupling limit, it has been demonstrated to reliably capture essential non-Markovian characteristics and intermediate coupling dynamics in structured environments\cite{zhao2025,10.1063/1.4942867}. In the present model, the system-environment interaction strength is selected within a regime where the Redfield equation remains numerically stable and physically consistent. In particular, we have explicitly verified that the reduced density matrix \(\rho(t)\)  retains positive semi-definiteness throughout the evolution under all parameter configurations considered.

Although alternative approaches such as the polaron-transformed master equation\cite{PhysRevE.89.042147} or hierarchical equations of motion (HEOM)\cite{PhysRevA.41.6676,Zhang2014NatComm} may provide advantages in specific strong-coupling regimes, the Redfield formalism offers a favorable balance between computational efficiency and descriptive accuracy. It thus serves as an appropriate framework for analyzing the energy transfer and ergotropy dynamics of the N-spin chain quantum battery investigated in this work. Its explicit form is given by Eq.(\ref{eq5}),

\vskip -0.1cm\begin{equation}
\mathcal{R}_L^{(\omega_{c})}[\rho] = \gamma(\omega_{c})(\hat{L} \rho \hat{L}^\dagger - \frac{1}{2} \{ \hat{L}^\dagger\hat{ L}, \hat{\rho} \}), \label{eq5}
\end{equation}

In the above expression, \(\gamma(\omega_{c})\) is a decay rate function derived from the environmental spectral density at frequency \(\omega_{c}\), which characterizes the strength of dissipation at this transition. The full dissipator \(\mathcal{R}[\rho]\) in Eq.(\ref{eq5}) is governed by the complete Redfield tensor, incorporating contributions from all relevant system-environment interactions. In the present scheme, we adopt the Debye model to describe the spectral density of the environment\cite{10.1063/1.4942867}, and the Bose-Einstein distribution factor to account for the thermal dissipation behavior during the charging process, shown in Eq.(\ref{eq6}).

\vskip -0.1cm\begin{equation}
\gamma(\omega_{c}) = \frac{J(\omega_{c})}{e^{\omega_{c} / (k_B T)} - 1}, \quad
J(\omega_{c}) = \kappa \cdot \frac{|\omega_{c}|}{\omega_a^2 + \omega_{c}^2}.    \label{eq6}
\end{equation}

\noindent In the Eq.(\ref{eq6}), \(\kappa \) denotes the cavity loss rate of the microcavity. In the present work, the temperature parameter \(T\) appearing in Eq.(\ref{eq6}) serves to characterize the thermal properties of the environment via the Bose-Einstein distribution in the spectral density. Throughout all simulations, we fix \(T\)=1 (in dimensionless units) to ensure a uniform thermal background across different parameter configurations. This allows us to focus on the influence of system-environment coupling parameters and external controls on the ergotropy dynamics. Although \(T\) variations may influence coherence and energy exchange especially when thermal decoherence is significant, while a detailed analysis of these effects may lie beyond the scope of this study.

If the initial state of the N-spin chain QB is set to the ground state\(|g\rangle_{b}\) and the initial state of the microcavity is the vacuum state \(|0\rangle_{c}\), then the initial state of the whole system is as follows:

\vskip -0.1cm\begin{equation}
|\psi_{0}\rangle=|g\rangle_{b} \bigotimes |0\rangle_{c}.    \label{eq7}
\end{equation}

\noindent which are the Kronecker products of two \(2^N\)-dimensional square matrices. The convolution in Eq.(\ref{eq4}) renders the dynamics non-Markovian. The energy storage in the QB at time \(t\) is given by

\vskip -0.1cm\begin{equation}
E_b(t) = \mathrm{Tr}\left[ \hat{H}_b \hat{\rho}_b(t) \right],   \label{eq8}
\end{equation}

\noindent where \(\rho_b(t)\) is the reduced density matrix of the QB at time \(t\). The stored energy in the QB is quantified as \(E_{b}(t)-E_{b}(0)\), where \(E_{b}(0)\)=\(E_{g}\) denotes the ground-state energy of the QB. Thus, the net charging energy is defined as

\vskip -0.1cm\begin{equation}
\Delta E_b(t) =E_b(t)-E_{g},     \label{eq9}
\end{equation}

\noindent which directly characterizes the energy transferred to the QB during the charging process. According to the second law of thermodynamics, no thermodynamic process can convert energy entirely into useful work without generating dissipative heat in a cyclic evolution. However, in a non-cyclic process-such as a closed-system unitary transformation from a non-passive state to its corresponding passive state-it is, in principle, possible to extract the entire charging energy \(\Delta E_b(t)\) in the form of work without any heat dissipation. In the present work, the thermal charging process is described by the Redfield master equation, which captures the dissipative and memory effects arising from the system-environment interaction. As a result, the energy transferred to the QB, \(\Delta E_b(t)\), generally exceeds the amount of useful work that can be extracted. The maximum extractable work, known as ergotropy, is then defined as follows \cite{Allahverdyan2004MaximalWork,PhysRevE.87.042123},

\vskip -0.1cm\begin{equation}
 \varepsilon_b(t) = E_b(t) - \min_U \operatorname{Tr}\left[ \hat{H}_b \hat{U} \hat{\rho}_b(t) \hat{U}^\dagger \right],     \label{eq10}
\end{equation}

\noindent where the eigenvalues of \(\hat{\rho}_b(t)\) are arranged in descending order, while the eigenvalues of \(\hat{H}_b\) are arranged in ascending order\cite{PhysRevB.99.035421}. Since the ergotropy is strictly bounded by \(\Delta E_b(t)\), the charging energy \(\Delta E_b(t)\) provides a reliable approximation for characterizing the performance of the QB.

We would like to point out that the exact expression of ergotropy, as originally formulated in Ref.\cite{Allahverdyan2004MaximalWork}, enables analytical evaluation by directly utilizing the eigenvalues and eigenstates of the density matrix and system Hamiltonian, without requiring explicit minimization. For the finite system sizes considered in this work (e.g., N = 2 or 3), such exact computation is indeed feasible. Nevertheless, in this study, we employ the net charging energy  \(\Delta E_b(t)\) as an upper bound to the ergotropy in order to efficiently capture the time-dependent energy dynamics within the non-Markovian framework described by the Redfield master equation. As the system size increases, calculating exact ergotropy becomes computationally demanding due to the exponential growth of the Hilbert space. Therefore,  \(\Delta E_b(t)\) serves as a practical and informative approximation for assessing the extractable energy. A more detailed comparison between  \(\Delta E_b(t)\) and the exact ergotropy \(\varepsilon_b(t)\), including an evaluation of the bound's tightness across different regimes, is an interesting direction for future investigation.

\section{Results and discussion}\label{discussion}
\subsection{Charging energy in Non-Markovian thermal charging process}

This section focuses on the charging properties of the QB through the net energy \(\Delta E_b(t)\), which serves as an approximate substitute for the ergotropy \(\varepsilon_b(t)\) in characterizing the extractable energy of the N-spin chain QB system.

\vskip 0.1cm
\begin{table}[htbp]
\caption{ Parameters used in this work. }
\begin{center}
\setlength{\tabcolsep}{0.1cm}\renewcommand{\arraystretch}{1.5} 
\resizebox{0.95\columnwidth}{!}{
\begin{tabular}{|c| c| c| c| c| c| c| c| c| c|}
\hline
Parameters & \multicolumn{3}{c|}{Fig.2} & \multicolumn{2}{c|}{Fig.3}  & \multicolumn{4}{c|}{Fig.4} \\
\hline
           & Fig.2(a)&Fig.2(b) &Fig.2(c) &Fig.3(a)&Fig.3(b) &Fig.4(a)&Fig.4(b)&Fig.4(c)&Fig.4(d) \\
\hline
$\omega_{c}$& 0.5    &$\backslash$& 0.35        & 0.5      & 0.5      & 0.5    & 0.5   & 0.5   & 0.5   \\
\hline
$\omega_{a}$&  1     & 0.25       &$\backslash$ & 1        &  1       & 1      & 1     &1      & 1            \\
\hline
$\omega_{d}$&$\pi$   & $1.25\pi$  & $1.25\pi$   &$0.75\pi$ & $1.05\pi$   &$\backslash$&3$\pi$ &$0.75\pi$& $\pi$ \\
\hline
$g$         &$\backslash$& 1.2    & 1.2         & 0.8      & 0.8     & 0.8     & 0.8   & 0.5   & 0.8     \\
\hline
$J$         & 0.95       & 0.95   & 0.95        &  0.95    & 0.95    & 0.95    & 0.95   &$\backslash$  &0.95    \\
\hline
$\kappa$    & 0.8        & 0.8    & 0.8         &  0.8     & 0.8     & 0.8     &  0.8   & 0.6  &$\backslash$ \\
\hline
$N$         &  2         & 2      & 2           &$\backslash$ &2     & 3       & 3      & 3    &2 \\
\hline
$n$         &  2         &2       & 2          &2          & $\backslash$ & 4  & 4      & 3    &2 \\
\hline
$A$         &  3         & 3     & 3           & 3         & 3           & 3   &$\backslash$ & 3  & 3      \\
\hline
\end{tabular}}
\label{Tab1}
\end{center}
\end{table}

By solving Eqs.(\ref{eq4}) and (\ref{eq9}), the charging dynamics of the N-spin coupled QB can be obtained. Before proceeding with quantitative analysis, it is necessary to predefine the parameters required for the proposed model. All relevant parameters used in the scheme are listed in Tab.(\ref{Tab1}). To facilitate physical interpretation,  Tab.(\ref{Tab1}) summarizes the parameter values employed across different figures, covering variations in cavity frequency \(\omega_c\), spin frequency \(\omega_a\), coupling strength \(g\), hopping rate \(J\), driving amplitude \(A\), and dissipation rate \(\kappa\). These parameters are chosen to lie within the regime of validity of the Redfield master equation, corresponding to moderate system-bath coupling and avoiding the ultra-strong coupling limit. Numerical checks confirm that the reduced density matrix remains positive semi-definite throughout the evolution, ensuring the physical consistency of the simulations. This framework enables a systematic exploration of how each parameter influences ergotropy dynamics while preserving the reliability of the underlying formalism.

\begin{figure}[h]
\begin{flushleft}
\includegraphics[width=0.5\columnwidth]{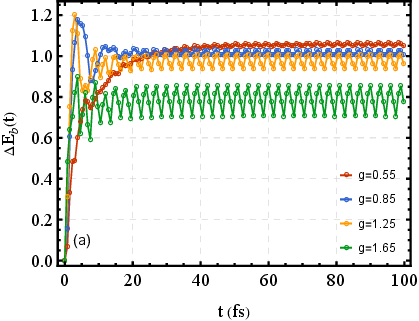}\includegraphics[width=0.5\columnwidth]{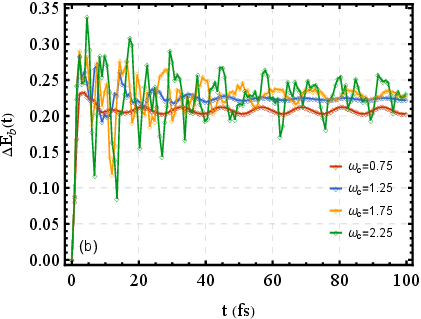}
\includegraphics[width=0.5\columnwidth]{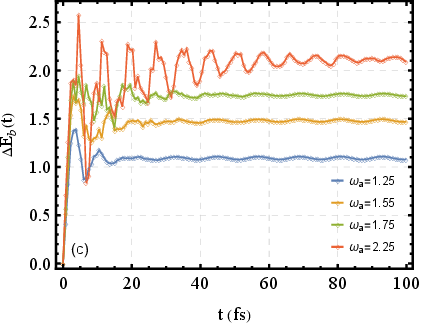}
\end{flushleft}
\caption{Oscillatory dynamics output of the charging energy under different (a). coupling strengths \(g\), (b). cavity field oscillation frequencies \(\omega_{c}\), and (c). the single-spin flip frequencies \(\omega_{a}\). Other parameters are taken from the Fig.2 column of Tab.~\ref{Tab1}.} \label{Fig2}
\end{figure}

Under the Markovian approximation, the output of the QB is typically obtained in a non-oscillatory form \cite{PhysRevA.103.033715}. However, since our proposed QB model takes into account strong coupling between the battery system and its surrounding environment, the net energy output of the battery exhibits oscillatory behavior during the thermal charging process. As shown in Fig.\ref{Fig2}(a), as the coupling coefficient \(g\) increases, the net extracted energy of the battery system gradually exhibits a stable oscillatory output. However, we also note that an excessively large \(g\) leads to a reduction in the net extracted energy output. As illustrated in Fig.\ref{Fig2}(b), when implementing cavity field frequency modulation through adjustments of the cavity parameters, the net extracted energy \(\Delta E_b(t)\) exhibits a slight variation in magnitude while demonstrating enhanced stability with intensified oscillatory characteristics. In all calculations, the parameter \(\omega_{a}\) which governs the total energy of the spin QB system is treated as a dimensionless quantity. As demonstrated in Fig.\ref{Fig2}(c), increasing parameter \(\omega_{a}\) not only decelerates the convergence to stable oscillatory output but also markedly amplifies the oscillation amplitude.

\begin{figure}[h]
\centering
\includegraphics[width=1\columnwidth]{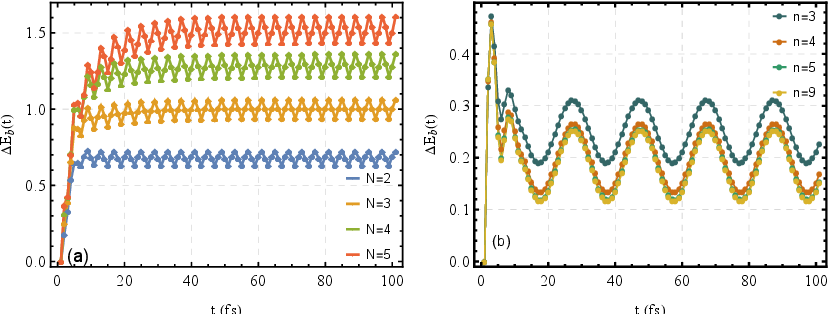}
\caption{Dynamical evolution of the charging energy under different numbers (a). the number \(N\) of spins in the QB, (b). the number \(n\) of photons in the cavity field. Other parameters are taken from the Fig.3 column of Tab.~\ref{Tab1}.}\label{Fig3}
\end{figure}

The number in N-spin chain QB emerges as a critical factor affecting net energy extraction \(\Delta E_b(t)\). As shown in Fig.\ref{Fig3}(a), increasing the spin number significantly enhances the maximum extractable work while marginally delaying output stabilization. In contrast, photon number in the microcavity induces slight amplitude attenuation of the net output energy without altering oscillation frequency, as evidenced in Fig.\ref{Fig3}(b).

\subsection{Suppression of oscillatory ergotropy.}

The oscillatory energy output characteristic remains incompatible with practical implementation requirements for QB systems. Therefore, suppressing such oscillatory outputs in our proposed quantum battery model constitutes a critical challenge requiring dedicated investigation. Notably, the present thermal charging protocol incorporates three tunable parameters: an external thermal coherence driving field, inter-spin hopping rate \(J\), and cavity dissipation factor \(\kappa\), which may provide viable control mechanisms for stabilizing these oscillatory outputs.

\begin{figure}[h]
\centering
\includegraphics[width=1\columnwidth]{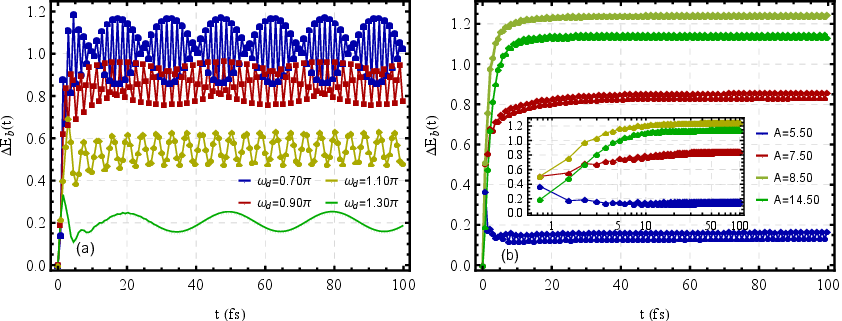}
\includegraphics[width=1\columnwidth]{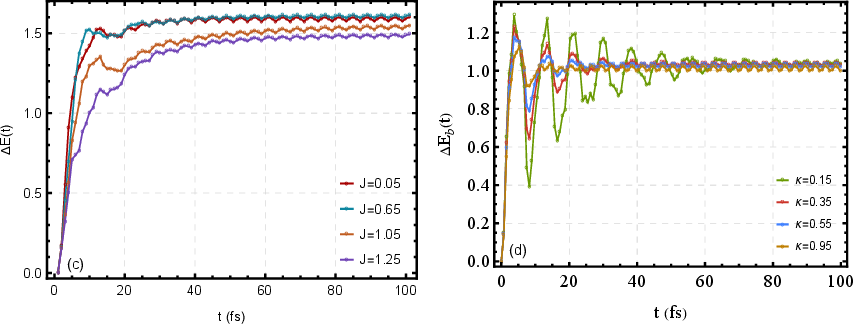}
\caption{Dynamical evolution of the charging energy with different parameters (a). external coherent driving field frequency \(\omega_{d}\) (b). external coherent driving field amplitude \(A\), (c). inter-spin hopping rate \(J\), (d). cavity dissipation factor \(\kappa\). Other parameters are taken from the Fig.4 column of Tab.~\ref{Tab1}.}\label{Fig4}
\end{figure}

Fig.\ref{Fig4}(a) demonstrates that the  frequency of the external field effectively suppresses oscillatory energy extraction, yet continuously increasing \(\omega_{d}\) significantly reduces net energy output, as evidenced by the contrasting cases of \(\omega_{d}\)=0.7\(\pi\) and  \(\omega_{d}\)=1.3\(\pi\) . A similar trend emerges in Fig.\ref{Fig4}(c) for the hopping rate \(J\): enhanced \(J\) values marginally diminish stable oscillatory outputs. In stark contrast, Fig.\ref{Fig4}(b) reveals that increasing the driving field amplitude substantially boosts net energy output; however, the case of \(A\) = 14.5 indicates an inhibitory effect at excessive amplitudes. Notably, cavity dissipation factor \(\kappa\) (Fig.\ref{Fig4}(d)) exhibits distinct behavior-its variation solely suppresses oscillations without altering the steady-state net energy value.

In particular, the results in Fig. 4(a) show that increasing the driving frequency \(\omega_{d}\) suppresses oscillations, though at the cost of reduced net energy gain. This can be interpreted as a detuning effect, where large \(\omega_{d}\) values weaken the resonance between the external field and the QB, reducing coherent energy transfer. Likewise, the inter-spin hopping rate \(J\) in Fig.\ref{Fig4}(c) modestly reduces oscillatory behavior, which can be ascribed to stronger intra-chain coherence that redistributes excitation among the spins and dampens local fluctuation.

In contrast, Fig.\ref{Fig4}(b) demonstrates that increasing the driving amplitude \(A\) enhances the energy inflow, strengthening the effective coupling between the drive and the QB; however, excessive amplitude (e.g.,\(A\)=14.5) leads to nonlinear saturation and eventual suppression. Finally, the observed suppression of oscillations in Fig.\ref{Fig4}(d) for large values of the cavity dissipation factor \(\kappa\) can be attributed to the dominance of the open-system dissipative contribution described by the Redfield tensor. As \(\kappa\) increases, the dissipative dynamics increasingly outweigh the unitary evolution governed by the system Hamiltonian. Since the coherent oscillations in ergotropy are primarily induced by the unitary term, this transition to dissipation-dominated dynamics naturally leads to diminished oscillations. This aligns with the intuitive understanding that decoherence suppresses coherent features of the system's evolution. Taken together, these results provide physical insight into how different parameters govern the interplay between coherent and dissipative processes, thereby enabling strategic suppression of oscillatory behavior and tunable control over ergotropy dynamics in our QB design.

\section{Conclusion}\label{conclusion}
In this work, we investigated the thermal charging dynamics of a non-Markovian QB composed of an N-spin chain coupled to a microcavity, employing the Redfield master equation to account for strong system-environment interactions. Our results reveal that, unlike the Markovian case, the ergotropy exhibits pronounced oscillatory output due to memory effects. These oscillations, however, can be effectively suppressed by tuning the coherent driving field, the cavity parameters, and the spin-spin hopping strength. Future work may incorporate multi-mode environments and realistic cavity-QED implementations, along with optimized control protocols to further improve charging efficiency and energy storage stability for practical applications.
\section{Acknowledgment}

This work is supported by the National Natural Science Foundation of China ( Grant Nos. 62065009 and 61565008 ),
Yunnan Fundamental Research Projects, China ( Grant No. 2016FB009 ) and the Foundation for Personnel training projects of Yunnan Province, China ( Grant No. KKSY201207068 ).

\section*{Data Availability Statement}

This manuscript has associated data in a data repository.[Authors' comment: All data included in this manuscript are available upon resonable request by contacting with the corresponding author.]

\section*{Conflict of Interest}
The authors declare that they have no conflict of interest. This article does not contain any studies with human participants or animals performed by any of the authors. Informed consent was obtained from all individual participants included in the study.

\bibliographystyle{apsrev4-1}
\bibliography{reference}
\end{document}